%% file: samplepaper.tex
\documentclass[runningheads]{llncs}
\usepackage{glossaries}
\makenoidxglossaries
\input{sections/glossary} 
\usepackage[T1]{fontenc}
\usepackage{graphicx}
\usepackage{adjustbox}
\begin{document}
\title{Micro-Collaborative Poisoning: A Distributed Attack on RAG Systems}
%
%
\author{Pedro Pereira\inst{1}\orcidID{0009-0008-7641-1566} \and
Eva Maia \inst{4}\orcidID{0000-0002-8075-531X} \and
Isabel Praça \inst{5}\orcidID{0000-0002-2519-9859}}

\authorrunning{P. Pereira et al.}

\institute{GECAD, ISEP, Polytechnic of Porto, Rua Dr. António Bernardino de Almeida, 4249-015 Porto, Portugal \\
\email{\{peesp,egm,icp\}@isep.ipp.pt}}
\maketitle              
\begin{abstract}
Retrieval-Augmented Generation (RAG) improves large language models by grounding outputs in external knowledge sources, but this dependency also creates a surface for poisoning attacks. This paper introduces Micro-Collaborative Poisoning, a distributed attack in which a false target claim is divided across multiple locally plausible documents instead of being concentrated in a single malicious passage. We evaluate the attack across 108 RAG configurations by varying dataset, retriever architecture, retrieval depth, database composition, number of poisoned databases, and generator model. The results indicate that Micro-Collaborative Poisoning is not driven by a single dominant poisoned passage, but by the accumulation of weak adversarial signals across retrieved sources. Increasing top-$k$ and poisoning multiple databases make it more likely that these signals will appear together in the retrieved context, while clean database diversity and stronger retrievers can reduce their influence. The document-level poisoning visibility analysis further shows that this threat is difficult to expose through isolated document inspection, since Micro-Collaborative Poisoning achieves downstream influence while leaving a weaker explicit poisoning signature than direct poisoning.

\keywords{Retrieval-Augmented Generation \and RAG Security \and Corpus Poisoning \and Knowledge Base Poisoning \and Distributed Poisoning}
\end{abstract}

\setlength{\textfloatsep}{9pt plus 2pt minus 2pt}
\section{Introduction}\label{sec:1}

\input{sections/Introduction}

\section{Related Work}\label{sec:2}
\input{sections/RelatedWork}

\section{Methodology}\label{sec:3}
\input{sections/Methods}

\section{Results and Discussion}\label{sec:4}

\input{sections/Results}

\section{Conclusions}\label{sec:5}

\input{sections/Conclusions}

\begin{credits}
\subsubsection{\ackname} This work was supported by the AIDA project, which has received funding from the European Defence Fund (EDF) under grant agreement 101168202. This work has also received funding from UID/00760/2025.
\end{credits}

\pagebreak
\bibliographystyle{splncs04}
\bibliography{sections/refs}
\end{document}

%% file: sections/glossary.tex
\newacronym{AI}{AI}{Artificial Intelligence}
\newacronym{NLP}{NLP}{Natural Language Processing}
\newacronym{LLMs}{LLMs}{Large Language Models}
\newacronym{RAG}{RAG}{Retrieval-Augmented Generation}
\newacronym{REALM}{REALM}{Retrieval-Augmented Language Model Pre-Training}
\newacronym{DPR}{DPR}{Dense Passage Retrieval}
\newacronym{CAR}{CAR}{Correct Answer Rate}
\newacronym{DPAR}{DPAR}{Document Poisoning Alignment Rate}
\newacronym{ASR}{ASR}{Attack Success Rate}
\newacronym{OLS}{OLS}{Ordinary Least Squares}

%% file: sections/Introduction.tex
\gls{RAG} has become one of the most important approaches to improving the practical reliability of \gls{LLMs} \cite{Huang}. By combining generation with external retrieval, \gls{RAG} allows a model to answer using information that is more recent, domain-specific, and grounded than the knowledge stored only in its parameters \cite{Lewis2020RetrievalAugmentedGF}. This makes it especially relevant for applications where factual accuracy, traceability, and adaptability are important, such as question answering, technical assistance, enterprise search, and decision support. Instead of treating the language model as a closed knowledge source, \gls{RAG} connects it to external repositories that can be updated, inspected, and adapted to specific use cases.

However, this same dependency on external knowledge also creates a new security risk. The final answer produced by a \gls{RAG} system is strongly influenced by the documents selected during retrieval, which means that corrupted or misleading documents can directly affect the generated response \cite{Zou_2025}. Prior work has shown that \gls{RAG} systems can be attacked through different parts of the pipeline. Some attacks manipulate the knowledge base directly by inserting false or misleading content \cite{Zou_2025}, while others target the retrieval stage by making poisoned documents more likely to be selected \cite{zhong-etal-2023-poisoning}. Additional strategies include backdoor attacks, where specific triggers cause the system to produce attacker-controlled outputs \cite{bagwe-etal-2025-rag}, and jamming attacks, where irrelevant or noisy content degrades the quality of retrieval and generation \cite{shafran2025machine}. 

Together, these attack types confirm that the external corpus is a critical security boundary for RAG systems. However, most of this work focuses on whether an attack succeeds, rather than whether the poisoned content can be detected once it is introduced. This distinction matters because a successful attack that is easily identified and filtered poses a far smaller practical threat than one capable of evading detection. More recent work has therefore studied stealthier poisoning strategies, showing that adversarial passages can be made more natural, coherent, and difficult to filter while still influencing the final answer \cite{Hidden}. Despite this progress, most existing approaches still place the adversarial signal inside one document or a small number of clearly attack-oriented passages. A less explored threat arises when no individual document appears overtly malicious, but several documents contain subtle and coordinated anomalies that collectively guide the model toward a false claim. This work studies this setting through Micro-Collaborative Poisoning, an attack in which a target misinformation claim is distributed across multiple locally plausible documents that appear benign in isolation but become influential when retrieved together.

The present work evaluates whether this type of low-detectability poisoning can exploit larger top-$K$ retrieval settings, where more documents are included in the generation context. The study compares this attack with the poisoning strategy analyzed in a previous study \cite{pereira2026influence} and measures its behavior under different retrieval and generation conditions, including dataset, retriever type, retrieval depth, number of databases, number of poisoned databases, and generator model. The analysis considers both retrieval-level and generation-level metrics, as well as the poisoning percentage of the generated documents, in order to understand not only whether the attack succeeds, but also how strongly and how visibly the poisoned signal is introduced.

The remainder of this paper is organized as follows. Section~\ref{sec:2} reviews the relevant background on \gls{RAG} and poisoning attacks. Section~\ref{sec:3} defines the Micro-Collaborative Poisoning attack, describing how the poisoned documents are generated, and presents the experimental setup, including datasets, retrieval models, database configurations, generator models, and evaluation metrics. Section~\ref{sec:4} discusses the results and compares Micro-Collaborative Poisoning with the previous poisoning approach. Finally, Section~\ref{sec:5} concludes the paper and outlines future directions for improving the security of \gls{RAG} systems.

%% file: sections/RelatedWork.tex
\gls{RAG} was introduced as a way to combine the generative capabilities of pre-trained language models with external, non-parametric knowledge sources \cite{Lewis2020RetrievalAugmentedGF}. Instead of relying only on information stored in model parameters, a \gls{RAG} pipeline retrieves relevant passages from an external corpus and grounds the generated response in this retrieved information \cite{guu2020realm,karpukhin2020dense}. 

However, the same mechanism that makes \gls{RAG} useful also makes it sensitive to the quality of retrieved context. If irrelevant, misleading, or adversarial passages are retrieved, the language model may incorporate them into the final answer. Prior work on retrieval robustness has shown that retrieval augmentation can sometimes reduce answer accuracy when the retrieved evidence is not useful or conflicts with the correct answer \cite{yoran2024making}. This motivates the security problem studied in this paper: once external documents become part of the model's reasoning process, the retrieval corpus itself becomes an attack surface.

Several works have shown that adversaries can exploit this attack surface by manipulating the documents available to the retriever. Early corpus poisoning work on dense retrieval demonstrated that small sets of adversarial passages can be inserted into a retrieval corpus to increase their likelihood of being retrieved for unseen queries \cite{zhong-etal-2023-poisoning}. Poisoning attacks against \gls{RAG} systems have therefore shown that injecting malicious texts into the knowledge base can cause the model to generate specific attacker chosen answers \cite{Zou_2025}.

Beyond direct corpus injection, other attacks target different components of the \gls{RAG} pipeline. Low-level perturbation attacks show that minor textual changes, such as noisy or typo modifications, can disrupt, thereby influencing influencing the final generated answer \cite{cho-etal-2024-typos}. Other attacks, such as jamming, pursue a different objective. Instead of forcing an incorrect answer, they insert blocker documents that cause the system to abstain or fail to answer a target query \cite{shafran2025machine}. Backdoor attacks further demonstrate that poisoned retrieval components or poisoned documents can create persistent trigger response behaviors, including fairness related manipulation where specific groups are associated with biased outputs \cite{bagwe-etal-2025-rag}.

More recent work has moved beyond measuring attack success alone and has begun to study whether poisoned content can remain difficult to detect. This is important because highly effective poisoned documents may also be easy to identify when they contain unnatural phrasing, excessive repetition, or explicit adversarial claims. Chang et al. show that a single poisoned document can remain effective for complex multi-hop questions when it is constructed as a credible, self-contained evidence chain with authority signals, achieving stronger attack performance while improving stealth against defenses \cite{chang-etal-2025-one}. Other studies operationalize stealth through a reinforcement-learning framework that iteratively applies small synonym-based perturbations to a target document, using retrieval, generation, and naturalness rewards to make the document enter the top-$k$ and then influence the generated answer while preserving semantic consistency and avoiding obvious surface-level anomalies \cite{song-etal-2025-silent}. Li et al. extend this direction at a finer granularity using a lightweight attacker model to generate malicious passages and then iteratively search for token-level substitutions that preserve query similarity while reducing the likelihood that the reader produces the correct answer \cite{li-etal-2026-token}. Together, these studies shift the focus from whether poisoning can succeed to how adversarial passages can be made fluent, credible, and difficult to distinguish from benign text.

Surface-level filters based on perplexity, repetition, or linguistic incoherence are insufficient against carefully constructed poisoned documents. Building on this line of work, we propose Micro-Collaborative Poisoning, a novel complementary threat model not previously explored in the literature. Rather than making a single poisoned document strong enough to dominate the retrieved context, the attack distributes the adversarial signal across several locally plausible documents. Following the configuration-based methodology of prior work on RAG poisoning influence factors, which showed that attack success depends on the interaction between retrieval configuration, database composition, and generator behavior, this study evaluates not only whether poisoning succeeds, but also under which pipeline conditions distributed weak signals become retrievable, mutually reinforcing, and influential during generation \cite{pereira2026influence}.

%% file: sections/Methods.tex
The experimental methodology is designed to analyze how poisoning behavior changes across different \gls{RAG} configurations. For this purpose, this section first presents the proposed Micro-Collaborative Poisoning attack and explains how adversarial documents are generated. It then describes the \gls{RAG} pipeline characteristics varied in the experiments, including the dataset, retriever architecture, retrieval depth, knowledge-base configuration, and generator model, resulting in a total of 108 experimental configurations. Finally, the section defines the evaluation metrics used to measure the attack from three perspectives: retrieval exposure, generation impact, and document-level poisoning visibility.

\subsection{Micro-Collaborative Poisoning Attack}

The Micro-Collaborative Poisoning attack is defined as a corpus-level poisoning strategy that inserts multiple adversarial documents into the knowledge bases used by the \gls{RAG} system. For each evaluation question, the attack receives as input the question $q$ and the expected correct answer $a$. The first step is to define an incorrect target claim $a_t$, which represents the answer that the attacker wants the system to generate. When the question contains an explicit alternative structure, such as ``A or B'', the target claim is selected as the incorrect alternative. Otherwise, the language model meta-llama-3-8b-instruct \cite{noeljacob_llama3_gguf} is used to generate plausible distractor answers, and the final target is selected after filtering out duplicates, semantically close variants of the correct answer, and candidates that are unrelated to the question.

After the target claim is selected, the attack generates a set of poisoned documents associated with the same question and target, using the same language model. Each document is produced as a short, single-paragraph passage written in a neutral reference style. The generated text must remain topically related to the original question, avoid mentioning the correct answer, and introduce support for the incorrect target claim. To preserve retrieval relevance, the generation process extracts anchor terms from the question. These terms correspond to the main entities, concepts, and keywords in $q$, after normalization, stop-word removal, and duplicate filtering. The anchor terms are then incorporated into the poisoned documents so that they remain lexically and semantically connected to the original query.

The attack controls poisoning visibility through a benign ratio, which specifies the proportion of each paragraph that should remain non-adversarial. The remaining proportion defines the poisoning ratio, the fraction of the paragraph available for the adversarial insertion. The method assumes a target paragraph length of 120 words and computes an insertion center by multiplying this length by the poisoning ratio. This center represents the expected insertion length before tolerance is applied. A 20\% tolerance is then applied around the center to obtain the insertion bounds. The minimum length is the lower tolerance value, constrained to be at least 6 words, while the maximum length is the upper tolerance value, constrained to be at most 50 words and at least two words larger than the minimum. Therefore, higher benign ratios force the adversarial signal to remain shorter and more locally plausible. By default, a benign ratio of 0.80 gives a poisoning ratio of 0.20, so the insertion center is ($0.20 \times 120 = 24$) words. Applying the tolerance gives an allowed range of approximately 19 to 29 words.

To distribute the adversarial influence, the attack generates five poisoned documents, assigning each one a unique rhetorical role. The roles used in the attack are (1) selective framing, which emphasizes contextual details that make the target claim appear more plausible, (2) semantic bridging, which connects the original question to concepts associated with the target claim, (3) boundary blurring, which introduces ambiguity between the correct answer and the incorrect target, (4) presupposition, which presents the target direction as background information rather than as a direct claim, and (5) light claim insertion, which provides the most explicit support for the target while remaining short and neutral. Together, these roles create a set of documents that support the same target claim through different forms of evidence.

Each generated document is validated before being inserted into the knowledge base. The validation process checks whether the document is a single coherent paragraph, contains the required anchor terms, supports the assigned rhetorical role, includes the target claim when required, and excludes the expected correct answer. It also checks whether the generated text contains hedging or contradictory language that could weaken the poisoning effect. If a document fails these checks, it is regenerated or repaired before being used in the experiment. Once validated, the poisoned documents are inserted into the database.

\subsection{Experimental Configuration}

Following Pereira et al. study \cite{pereira2026influence}, the evaluation varies the main components of the \gls{RAG} pipeline in order to measure how each factor affects poisoning behavior. The first factor is the dataset. Experiments are conducted using HotpotQA \cite{yang-etal-2018-hotpotqa} and MS-MARCO \cite{nguyen2016ms}. HotpotQA represents a multi-hop question-answering setting where evidence may be distributed across several documents, while MS-MARCO represents open-domain passage retrieval over web-style content. To make the comparison computationally feasible and controlled between the two datasets, the experiments use a curated subset of 100 semantically aligned question pairs between the two datasets.

The second factor is the retriever architecture. Three retrieval methods are evaluated: BM25, dense BGE, and graph-based retrieval. BM25 represents sparse lexical retrieval and ranks documents according to term-frequency and inverse-document-frequency statistics. Dense BGE represents semantic vector retrieval, where queries and documents are encoded into dense embeddings and compared in vector space. Graph-based retrieval incorporates structural relationships between documents and retrieves evidence by considering connections between related pieces of information. These three architectures allow the experiments to compare lexical, semantic, and relational retrieval behavior under poisoning.

The third factor is retrieval depth, controlled through the top-$k$ parameter. The experiments evaluate top-$k$ values of 2, 3, and 5. This parameter is particularly important for Micro-Collaborative Poisoning because the attack depends on the possibility that multiple weak poisoned documents are retrieved together. Smaller values of $k$ restrict the context and may prevent the collaborative signal from accumulating. Larger values provide more evidence to the generator, but also increase the probability that several poisoned documents enter the context at the same time.

The fourth factor is the knowledge-base configuration. The experiments consider systems with either one or two databases. In the two-database setting, the same dataset is replicated across both sources to simulate a redundant retrieval environment. The number of poisoned databases is then varied to represent different compromise levels. Three configurations are evaluated: one database with one poisoned source, two databases with one poisoned source, and two databases with two poisoned sources. These configurations allow the study to compare a single-source compromise, a partially compromised redundant system, and a fully compromised multi-source setting.

The fifth factor is the generator model. Two \gls{LLMs} are evaluated: llama-4-scout-17b-16e-instruct and openai-gpt-oss-120b. Since the generator is responsible for interpreting the retrieved context and producing the final answer, comparing different models makes it possible to analyze whether poisoning success depends only on retrieval exposure or also on model-specific generation behavior, such as the tendency to follow misleading evidence, preserve the correct answer, or abstain.

Combining the varied factors results in a full factorial design with 108 configurations. For each configuration, poisoned documents are inserted into the selected knowledge bases, the retriever returns the top-$k$ documents for each query, and the generator produces an answer using the retrieved context. At the end, the retrieved documents and final responses are evaluated using the metrics described below.

\subsection{Evaluation Metrics}

The evaluation measures the attack from three perspectives: retrieval metrics, generation metrics, and document-level poisoning visibility. For retrieval metrics, we calculate Poison@k, which shows whether a poisoned document is included among the top-$k$ retrieved results \cite{Zou_2025,chang-etal-2025-one,wang2026jointgcg}, Poison Rank, which records the highest position of any poisoned document \cite{wang2026jointgcg}, and Score Margin, which represents the difference in similarity score between the strongest poisoned document and the top-ranked clean document \cite{wang2026jointgcg}. At the generation stage, we examine the model’s final response when retrieval includes poisoned content. The generator may respond in one of three ways: it may refuse to answer if the evidence seems untrustworthy, captured by the Abstention Rate \cite{chen-etal-2024-controlling}; it may still provide the correct answer, measured by the \gls{CAR} \cite{Zou_2025,chang-etal-2025-one}; or it may be affected by the poisoned evidence and produce a manipulated response, measured by the \gls{ASR} \cite{Zou_2025,chang-etal-2025-one}.

Document-level poisoning visibility measures how explicit or detectable the poisoned signal is within adversarial documents. While retrieval and generation metrics show whether poisoned documents are retrieved and influence the final answer, they do not assess how visible adversarial intent is in each document. This is important for Micro-Collaborative Poisoning, which reduces the explicitness of individual poisoned documents so each may support the target claim without appearing malicious in isolation. Therefore, attack success should be evaluated alongside how much each document exposes its adversarial intent.
 
To address this, we introduce \gls{DPAR}, a document-level metric that estimates the poisoning visibility of generated adversarial documents. For each generated document, \gls{DPAR} receives four inputs: the original question $q$, the expected correct answer $a$, the adversarial target claim $t$, and the generated document $d_i$. The document is first divided into sentences:
\begin{equation}
  d_i = \left(s_{i,1}, s_{i,2}, \ldots, s_{i,n_i}\right)
  \label{eq:document-sentence-sequence}
\end{equation}

where $n_i$ is the number of sentences in document $d_i$, and $s_{i,t}$ denotes the $t^{th}$ sentence of the document $d_i$. Each sentence $s_{i,t}$ is then evaluated according to five poisoning indicators. Target-claim presence ($f_{\text{target}}(s_{i,t})$) measures whether the sentence contains the adversarial target claim and whether it is connected to the question. A direct target mention contributes 45 points, since it is the strongest poisoning signal.  An additional $35 \cdot \rho(s_i, q)$ points are added when the target claim co-occurs with semantically relevant role terms extracted from the question, such as person names, dates, or locations, where $\rho(s_i, q) \in [0,1]$ measures the overlap between $s_i$ and the question’s key role vocabulary. If the target claim and expected answer appear together, 10 points are added, as this may indicate contradiction or answer substitution.

Instruction-like language ($f_{\text{instr}}(s_{i,t})$) captures language patterns that try to guide the model toward a specific answer, such as directive or answer-focused wording like ``ignore'' or ``always say''. It gives a score of 0, 10 or 20, depending on whether zero, one, or multiple instruction patterns are found. 

Artificial authority ($f_{\text{auth}}(s_{i,t})$) detects ungrounded wording that presents the claim as official, updated, authoritative, or definitive, using expressions such as ``official'' or ``most recent''. It gives a score of 0, 5 or 10 depending on whether zero, one, or multiple authority cues are found. 

Keyword stuffing ($f_{\text{kw}}(s_{i,t})$) measures unnatural repetition of question terms intended to improve retrieval. Densities exceeding 20\% of the sentence's tokens contribute 4 points, densities exceeding 35\% contribute an additional 4 points, and the presence of three or more repeated role terms contributes a further 2 points. 

Contextual propagation ($f_{\text{ctx}}(s_{i,t})$) captures sentences that do not directly mention the target claim but reinforce a previous poisoned or suspicious sentence. A score of 20 points is awarded to sentence $s_i$ if the preceding sentence $s_{i-1}$ was itself classified as suspicious or contained the target claim $a_t$.

These indicators are combined into a sentence poison score $P(s_i)$, bounded between 0 and 100:
 \begin{equation}
 \small
  P(s_i)
  =
  \min\Bigl(
  f_{\text{target}}(s_{i,t})
  +
  f_{\text{instr}}(s_{i,t})
  +
  f_{\text{auth}}(s_{i,t})
  +
  f_{\text{kw}}(s_{i,t})
  +
  f_{\text{ctx}}(s_{i,t}),
  \ 100
  \Bigr).
  \label{eq:sentence-priority-score}
  \end{equation}

The point values are chosen to reflect the relative visibility of each poisoning signal. Direct target mentions receive the highest score because they explicitly expose the adversarial claim. Overlap with the question, instruction-like wording, and contextual propagation receive intermediate scores because they make the poisoned content more likely to influence the answer, and artificial authority or keyword stuffing receive lower scores because they mainly strengthen or expose the poisoned signal rather than directly stating it. The score is first allowed to accumulate beyond 100 so that multiple poisoning signals can reinforce each other within the same sentence. It is then capped at 100 to keep $P(s_i)$ on a fixed and comparable severity scale, preventing extreme combinations of indicators from dominating the document-level score.

After computing $P(s_i)$, each sentence is assigned a label according to its score and content. Sentences with $P(s_i) \geq 70$ are labeled \texttt{Poisoned}, whereas those with $40 \leq P(s_i) < 70$ are labeled \texttt{Suspicious}. If $P(s_i) < 40$ and the sentence contains the expected answer without containing the adversarial target claim $t$, it is labeled as \texttt{Normal Supporting}. All remaining low-scoring sentences are labeled as \texttt{Normal Background}. This segmentation step is important because a document may contain both benign and adversarial content. Instead of assigning a single binary label to the whole document, \gls{DPAR} identifies which parts of the document contribute to the poisoned signal and how strongly they do so.

The sentence scores are then aggregated into a document poisoning score, denoted by $\Delta$, which represents the overall poisoning visibility of the generated document. Let $n$ be the number of sentences in the document and let $P(s_i)$ be the poison score of sentence $s_i$. The aggregation combines three complementary properties: the proportion of poisoned sentences, $\pi_P=\frac{N_P}{n}$, where $N_P$ is the number of sentences with $P(s_i)\geq70$, the average sentence poison score, $\bar{P}=\frac{1}{n}\sum_{i=1}^{n}P(s_i)$, and the maximum sentence poison score, $P_{\max}=\max_i P(s_i)$. The final document-level score is computed as:

\begin{equation}
    \Delta \;=\; 0.45\;\pi_P \cdot 100
            \;+\; 0.35\;\bar{P}
            \;+\; 0.20\;P_{\max}.
  \label{eq:a}
\end{equation}

The document is labeled as \texttt{Normal} when $\Delta < 30$, as \texttt{Suspicious} when $30 \leq \Delta < 50$, as \texttt{Poisoning} when $50 \leq \Delta < 70$, and as \texttt{Strong Poisoning} when $\Delta \geq 70$. This classification then makes it possible to compare poisoning strategies not only by whether they succeed, but also by how visible or detectable their generated documents are.

%% file: sections/Results.tex
The experimental analysis starts by evaluating Micro-Collaborative Poisoning across both retrieval-level and generation-level metrics. We first examine the main effects of the experimental factors using \gls{OLS} regression models \cite{ols} fitted independently for each metric. The models are used as an explanatory analysis rather than prediction. Each coefficient estimates the average change associated with a configuration factor while controlling for the remaining factors in the experimental grid. Because the experiments vary several parameters simultaneously, the regression coefficients help isolate the relative influence of each parameter.

Table~\ref{tab:retrieval_main_effects_micro_poisoning} reports the strongest main effects for the retrieval-level metrics. For Poison@k, retrieval depth is the dominant factor. Increasing the number of retrieved documents from $k=2$ to $k=5$ produces the largest positive effect, while increasing to $k=3$ also substantially increases poisoning exposure. This confirms that broader retrieval windows increase the probability that at least one poisoned passage enters the context. This effect is especially important for Micro-Collaborative Poisoning, because the attack benefits from the accumulation of multiple weak adversarial signals. Dataset choice also has a clear effect on Poison@k. Experiments on MS-MARCO reduce the probability of retrieving poisoned content when compared with HotpotQA. This suggests that dataset structure affects poisoning susceptibility. HotpotQA often involves multi-hop reasoning and evidence aggregation, which can make it easier for distributed poisoned passages to appear relevant to the query. MS-MARCO, by contrast, appears to provide stronger competition from clean passages or a retrieval structure in which poisoned passages are less frequently selected. The graph-based retriever also reduces Poison@k, showing that structure-aware retrieval can still limit exposure to adversarial content. 

\begin{table}[h]
  \centering
  \caption{Top 5 strongest main effects for retrieval-level metrics under Micro-Collaborative Poisoning.}
  \label{tab:retrieval_main_effects_micro_poisoning}
  \resizebox{\textwidth}{!}{%
  \begin{tabular}{llllll}
  \hline
  \multicolumn{2}{c}{\textbf{Poison@k}} &
  \multicolumn{2}{c}{\textbf{Poison Rank}} &
  \multicolumn{2}{c}{\textbf{Score Margin}} \\
  \hline
  Factor & Coef. & Factor & Coef. & Factor & Coef. \\
  \hline
  C(top\_k)[T.5] & 0.141 & C(num\_databases)[T.2] & 1.714 & C(top\_k)[T.5] & 0.055 \\
  C(dataset)[T.MS-MARCO] & -0.090 & C(num\_poisoned\_databases)[T.2] & -1.602 & C(retriever)[T.dense\_bge] & -0.050 \\
  C(top\_k)[T.3] & 0.082 & C(top\_k)[T.5] & 0.905 & C(retriever)[T.Graph] & -0.041 \\
  C(retriever)[T.Graph] & -0.068 & C(dataset)[T.MS-MARCO] & 0.345 & C(dataset)[T.MS-MARCO] & -0.038 \\
  C(num\_poisoned\_databases)[T.2] & 0.035 & C(top\_k)[T.3] & 0.322 & C(num\_poisoned\_databases)[T.2] & 0.038 \\
  \hline
  \end{tabular}%
  }
\end{table}

The Poison Rank results show a complementary pattern. Increasing the total number of databases raises Poison Rank, meaning that poisoned passages tend to appear lower in the ranking when they compete against a broader clean evidence pool. This is a robustness-enhancing effect, adding an additional database introduces more clean candidates, which can push poisoned passages away from the top positions. In contrast, increasing the number of poisoned databases strongly decreases Poison Rank. When poisoned content is present in more than one database, adversarial passages have more opportunities to match the query and occupy high-ranking positions. This demonstrates that database diversity is beneficial only when it increases the amount of clean competing evidence. If the additional database is also poisoned, the same multi-source setting becomes an advantage for the attacker.

The Score Margin results further clarify how competitive poisoned passages are once they are retrieved. Both dense BGE and graph retrieval reduce the Score Margin, meaning that poisoned passages are less strongly separated from clean evidence under these retrieval strategies. MS-MARCO also reduces the Score Margin, reinforcing the conclusion that this dataset is less susceptible than HotpotQA. However, poisoning two databases increases the Score Margin. This indicates that adversarial redundancy across databases makes poisoned passages more competitive relative to clean documents. Taken together, the retrieval-level results show that Micro-Collaborative Poisoning is shaped by a balance between retrieval depth, clean-source competition, and adversarial coverage. Larger retrieval windows and multiple poisoned databases increase risk, while clean database diversity and robust retrieval architectures reduce it.

We next examine generation-level metrics. Table~\ref{tab:generation_main_micro_poisoning} summarizes the strongest main effects for Abstention Rate, \gls{CAR}, and \gls{ASR}. For the Abstention Rate, the strongest effect is associated with the openai-gpt-oss-120b generator, which decreases abstention. This indicates that this model is more likely to produce an answer rather than refuse or withhold judgment when the retrieved context contains conflicting or misleading information. Retrieval depth also decreases abstention, especially at $k=5$. This suggests that providing more retrieved passages increases the model's tendency to answer, possibly because a larger context gives the appearance of greater evidential support. MS-MARCO also reduces abstention, while the graph-based retriever increases it. The graph effect may indicate that graph retrieval produces evidence sets that are less directly answerable or more cautious from the model's perspective, leading to a higher probability of abstention.

\begin{table}[h]
  \centering
  \caption{Top 5 strongest main effects for generation-level metrics under Micro-Collaborative Poisoning.}
  \label{tab:generation_main_micro_poisoning}
  \resizebox{\textwidth}{!}{%
  \begin{tabular}{llllll}
  \hline
  \multicolumn{2}{c}{\textbf{Abstention Rate}} &
  \multicolumn{2}{c}{\textbf{\gls{CAR}}} &
  \multicolumn{2}{c}{\textbf{ASR}} \\
  \hline
  Factor & Coef. & Factor & Coef. & Factor & Coef. \\
  \hline
  C(llm)[T.openai-gpt-oss-120b] & -0.064 & C(num\_databases)[T.2] & 0.124 & C(num\_poisoned\_databases)[T.2] & 0.110 \\
  C(top\_k)[T.5] & -0.056 & C(num\_poisoned\_databases)[T.2] & -0.116 & C(llm)[T.openai-gpt-oss-120b] & 0.088 \\
  C(dataset)[T.MS-MARCO] & -0.054 & C(retriever)[T.dense\_bge] & 0.045 & C(num\_databases)[T.2] & -0.084 \\
  C(retriever)[T.Graph] & 0.046 & C(llm)[T.openai-gpt-oss-120b] & -0.024 & C(top\_k)[T.5] & 0.076 \\
  C(num\_databases)[T.2] & -0.040 & C(top\_k)[T.3] & -0.021 & C(retriever)[T.Graph] & -0.064 \\
  \hline
  \end{tabular}%
  }
\end{table}

The \gls{CAR} results show that database composition has the strongest influence on answer correctness. Using two databases increases \gls{CAR}, suggesting that additional clean evidence improves the generator's ability to recover the correct answer. In contrast, poisoning two databases reduces \gls{CAR}. This confirms that distributed adversarial coverage can degrade answer quality even when each individual poisoned passage is only weakly manipulated. Dense BGE retrieval improves \gls{CAR}, indicating that semantic retrieval provides more useful evidence for answer generation. The openai-gpt-oss-120b model reduces \gls{CAR}, which is consistent with its lower abstention rate. The model answers more often, but this greater willingness to answer also increases the risk of producing an incorrect or adversarially influenced response.

The \gls{ASR} results provide the clearest view of end-to-end attack success. Poisoning two databases has the largest positive effect on \gls{ASR}, showing that adversarial coverage across multiple sources is the central driver of Micro-Collaborative Poisoning. The openai-gpt-oss-120b generator also increases \gls{ASR}, confirming that model behavior plays a major role after retrieval has occurred. By contrast, using two total databases reduces \gls{ASR}, again showing the protective effect of clean-source competition. Increasing retrieval depth to $k=5$ increases \gls{ASR}. This aligns with the Poison@k results, when more documents are retrieved, more poisoned fragments can enter the context and collectively influence generation. Finally, the graph-based retriever reduces \gls{ASR}, indicating that retrieval architecture can mitigate the downstream effect of poisoning, even when it does not fully eliminate poisoned passages from the retrieved set.

Overall, the main-effect analysis shows that Micro-Collaborative Poisoning is not controlled by a single factor. Instead, it emerges from the interaction between retrieval depth, database composition, retriever architecture, dataset characteristics, and generator behavior. The most important pattern is the contrast between total database diversity and poisoned database coverage. Adding clean databases improves robustness by increasing competition from legitimate evidence, while poisoning multiple databases strengthens the attack by making adversarial evidence more redundant and more likely to appear in the final context.

Because poisoning behavior in \gls{RAG} systems may depend on combinations of factors, we also analyze interaction effects using an extended \gls{OLS} model for \gls{ASR}. Unlike the main-effect models, this specification includes both the individual contribution of each factor and interaction terms up to the third order, allowing us to test whether the effect of one parameter changes depending on the value of others. Tables~\ref{tab:interaction_effects_asr_micro_poisoning} and~\ref{tab:threeway_effects_asr_micro_poisoning} summarize two complementary views of this model. Table~\ref{tab:interaction_effects_asr_micro_poisoning} reports the strongest terms in the full \gls{ASR} interaction model, ranked by the absolute value of their coefficients, regardless of whether they correspond to main effects, two-way interactions, or three-way interactions. In contrast, Table~\ref{tab:threeway_effects_asr_micro_poisoning} filters the same fitted model to include only pure up to three-way interaction terms, which capture cases where attack success depends on a specific combination of two or three experimental conditions. We focus on \gls{ASR} because it is the most direct end-to-end measure of attack effectiveness.

\vspace{-1.0em}
\begin{table}[h]
  \centering
  \caption{Top 5 strongest interaction effects on \gls{ASR} under Micro-Collaborative Poisoning.}
  \label{tab:interaction_effects_asr_micro_poisoning}
  \begin{adjustbox}{max width=0.8\textwidth}
  \begin{tabular}{ll}
  \hline
  \textbf{Factor / Interaction} & \textbf{Coef.}  \\
  \hline
  C(dataset)[T.MS-MARCO]:C(retriever)[T.dense\_bge]:C(top\_k)[T.5] & 0.142  \\
  C(num\_databases)[T.2] & -0.097  \\
  C(dataset)[T.MS-MARCO]:C(num\_poisoned\_databases)[T.2]:C(llm)[T.openai-gpt-oss-120b] & -0.092  \\
  C(dataset)[T.MS-MARCO]:C(retriever)[T.dense\_bge]:C(top\_k)[T.3] & 0.088  \\
  C(retriever)[T.Graph]:C(num\_poisoned\_databases)[T.2]:C(llm)[T.openai-gpt-oss-120b] & -0.079  \\
  \hline
  \end{tabular}%
  \end{adjustbox}
\end{table}
\vspace{-2.5em}

\begin{table}[h]
  \centering
  \caption{Top 5 strongest pure up to three-way interactions on \gls{ASR} under Micro-Collaborative Poisoning.}
  \label{tab:threeway_effects_asr_micro_poisoning}
  \begin{adjustbox}{max width=0.8\textwidth}
  \begin{tabular}{ll}
  \hline
  \textbf{Factor / Interaction} & \textbf{Coef.}  \\
  \hline
  C(dataset)[T.MS-MARCO]:C(retriever)[T.dense\_bge]:C(top\_k)[T.5]    & 0.142  \\
  C(num\_poisoned\_databases)[2]:C(llm)[T.openai-gpt-oss-120b]        & 0.104  \\
  C(retriever)[dense\_bge]:C(llm)[T.openai-gpt-oss-120b]             & 0.103  \\
  C(dataset)[T.MS-MARCO]:C(top\_k)[5]                                & 0.102  \\
  C(dataset)[HotpotQA]:C(llm)[T.openai-gpt-oss-120b]                & 0.099  \\
  \hline
  \end{tabular}%
  \end{adjustbox}
\end{table}
\vspace{-1.0em}

The interaction results show that dataset, retriever, and retrieval depth jointly shape attack success. The strongest interaction involves MS-MARCO, dense BGE, and $k=5$, with a positive effect on \gls{ASR}. This is important because dense retrieval appears protective in the main-effect analysis, yet under specific dataset and retrieval-depth conditions this advantage can be reduced. A similar positive interaction appears for MS-MARCO, dense BGE, and $k=3$. These findings show that robust retrieval methods cannot be evaluated only through their average effect. Their behavior depends on the dataset and on how much context is retrieved.

The interaction involving dataset, poisoned databases, and generator model has a negative effect for MS-MARCO with two poisoned databases and openai-gpt-oss-120b. This suggests that the effect of adversarial database coverage is not uniform across datasets and models. Some configurations reduce attack success despite the presence of multiple poisoned databases, likely because the retrieved evidence remains sufficiently competitive or because the model does not consistently resolve the context in favor of the poisoned target. Similarly, the interaction between graph retrieval, poisoned databases, and openai-gpt-oss-120b has a negative effect, reinforcing the idea that graph retrieval can reduce downstream attack success even in settings where the generator is otherwise vulnerable.

\begin{figure}[h]
\centering
\includegraphics[width=1\textwidth]{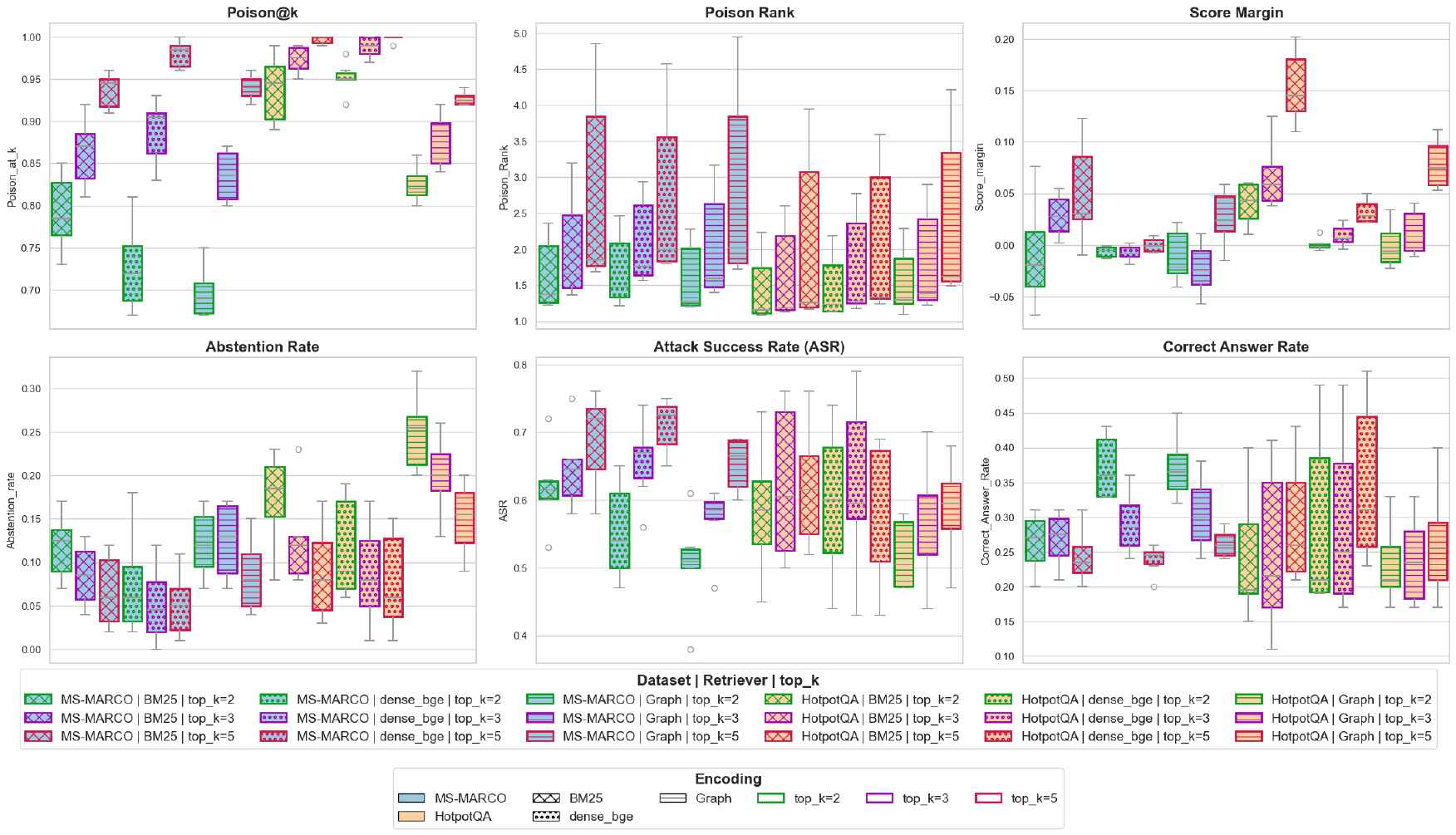}
\caption{Impact of the interaction between dataset, retriever, and retrieval depth on retrieval-level and generation-level metrics under Micro-Collaborative Poisoning.}
\label{fig:dataset_retriever_topk_micro_poisoning}
\end{figure}

Based on these findings, Fig.~\ref{fig:dataset_retriever_topk_micro_poisoning} illustrates with more detail the interaction between dataset, retriever, and retrieval depth across all retrieval and generation metrics. The figure confirms that Poison@k generally increases with retrieval depth. This trend is expected, since retrieving more documents increases the chance that poisoned passages enter the context. However, the magnitude of this increase depends strongly on the dataset and retriever. BM25 is generally the most exposed retriever, reflecting the vulnerability of lexical matching to adversarial passages that share surface-level terms with the query. Dense BGE and graph retrieval are more robust overall, but their advantage is not uniform across datasets, especially at high levels of k.

The Poison Rank and Score Margin panels provide further insight into this behavior. Under stronger retrieval settings, poisoned passages are often pushed lower in the ranking or become less competitive with clean evidence. This effect is clearer on MS-MARCO, where dense and graph retrieval reduce poisoning competitiveness more consistently. HotpotQA remains more vulnerable, which suggests that multi-hop or compositional evidence settings make it easier for poisoned passages to appear useful to the retriever. In such cases, even structure-aware retrieval may retrieve adversarial passages if they appear to contribute to one step of the reasoning chain.

The generation-level panels in Fig.~\ref{fig:dataset_retriever_topk_micro_poisoning} show how retrieval differences propagate to final answers. Higher Poison@k does not always translate directly into higher \gls{ASR}, because the generator may abstain or may still rely on clean evidence. Nevertheless, the general trend is clear, configurations with higher poisoning exposure tend to produce higher attack success, especially when the retrieval depth is larger and the retriever is less robust. Graph retrieval often reduces \gls{ASR}, particularly in settings where it also reduces poisoning competitiveness. Dense BGE also improves robustness in several configurations, although the interaction analysis shows that its protective effect can weaken under larger retrieval windows.

\begin{figure}[h]
\centering
\includegraphics[width=1\textwidth]{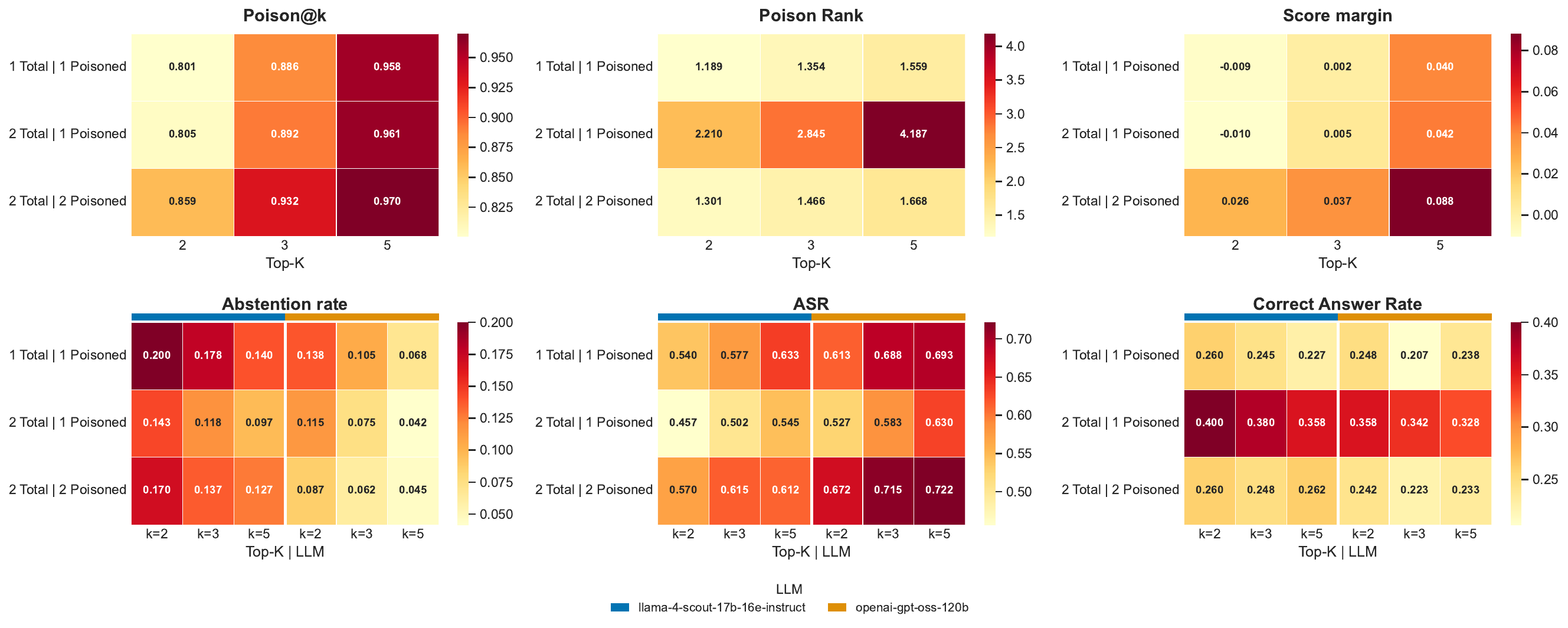}
\caption{Mean retrieval and generation metrics across database configurations under Micro-Collaborative Poisoning.}
\label{fig:db_config_llm_heatmaps_micro_poisoning}
\end{figure}

Fig.~\ref{fig:db_config_llm_heatmaps_micro_poisoning} presents an additional analysis focusing on database configuration, retrieval depth, and generator model. This figure is especially important for Micro-Collaborative Poisoning because the attack is explicitly designed around distributed evidence. At the retrieval level, Poison@k increases with retrieval depth across all database configurations. However, database composition mainly determines how competitive the poisoned passages are. The ``2 Databases $|$ 1 Poisoned'' setting tends to push poisoned passages lower in the ranking, because the poisoned database must compete with an additional clean source. This explains the positive effect of two total databases on Poison Rank and the negative effect on \gls{ASR}. In contrast, the ``2 Databases $|$ 2 Poisoned'' setting increases adversarial coverage. When both databases contain poisoned content, the retriever has more opportunities to select poisoned passages, and the generator is more likely to receive a context in which multiple documents support the same misleading claim.

\vspace{-.5em}
\vspace{-1.0em}

At the generation level, the heatmaps show clear differences between the evaluated LLMs. The openai-gpt-oss-120b model generally abstains less and exhibits higher \gls{ASR}, indicating that it is more willing to produce an answer and more likely to follow poisoned evidence. The llama-4-scout-17b-16e-instruct model is comparatively more cautious, with higher abstention and lower attack success in several settings. This contrast shows that generator-level robustness is not simply a function of retrieval quality. Even with the same retrieved evidence, different LLMs may differ in how they resolve conflicts, whether they defer judgment, and how strongly they rely on repeated evidence in the context.

The same figure also shows that increasing retrieval depth can amplify the effect of poisoning. At low retrieval depth, the context may contain too few poisoned passages for the collaborative signal to dominate. At larger $k$, however, multiple weakly poisoned passages can appear together. This is the central mechanism of Micro-Collaborative Poisoning, the attack does not require one obviously malicious passage to outrank all clean evidence. Instead, it relies on several plausible passages that collectively shift the generator toward the adversarial answer. This makes the attack more subtle than direct poisoning, because each individual poisoned passage may appear less suspicious.

\begin{figure}[h]
\centering
\includegraphics[width=0.7\textwidth]{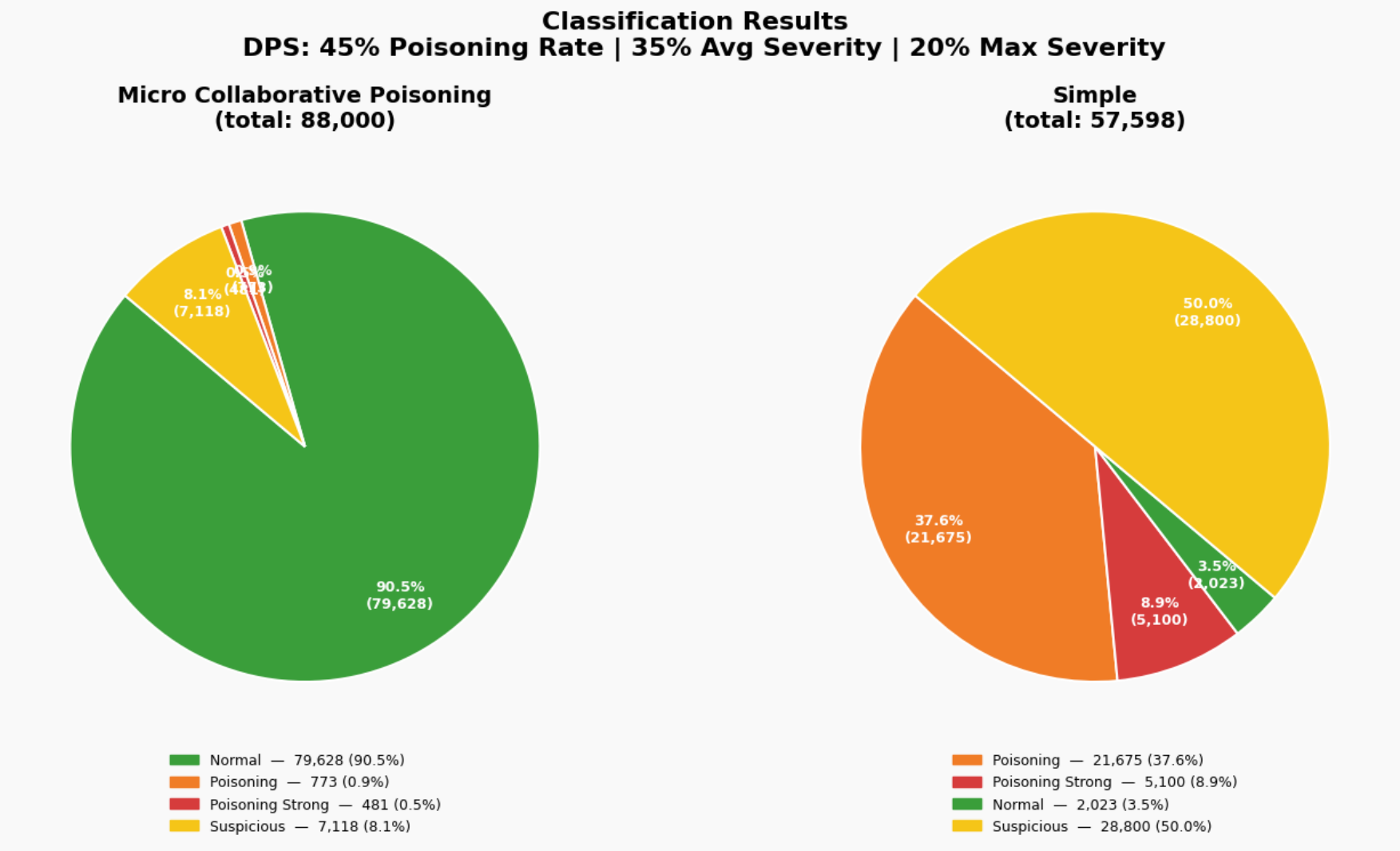}
\caption{Classification results comparing Micro-Collaborative Poisoning and the direct poisoning baseline under \gls{DPAR}.}
\label{fig:classification_results}
\end{figure}

Finally, Fig.~\ref{fig:classification_results} compares Micro-Collaborative Poisoning with the direct poisoning baseline \cite{pereira2026influence} under \gls{DPAR}. The results show a major difference in detectability. Under Micro-Collaborative Poisoning, most segments are classified as \textsc{Normal}, while only a small fraction fall into explicit poisoning categories. By contrast, the direct baseline produces far fewer \textsc{Normal} classifications and many more explicit poisoning detections. This confirms that Micro-Collaborative Poisoning has a weaker document-level signature. The attack is therefore not only a question of retrieval or generation performance, but also of stealth.

\vspace{-1.0em}

Compared with the direct poisoning setting reported by Pereira et al. \cite{pereira2026influence}, Micro-Collaborative Poisoning exhibits a different robustness profile. In the direct attack, retriever architecture is the dominant protective factor. The Graph retriever produces the strongest reduction in Poison@k and substantially lowers \gls{ASR}, while dense BGE also strongly reduces Score Margin and attack success. In contrast, under Micro-Collaborative Poisoning, retrieval depth and database composition become more influential. Increasing $k$ has a stronger effect on Poison@k, and poisoning two databases has the largest positive effect on \gls{ASR}, showing that the Micro-Collaborative strategy depends less on a single highly ranked poisoned document and more on the accumulation of several weak adversarial passages across the retrieved context. The protective effect of Graph and dense retrieval is still present, but it is weaker than in the direct poisoning case, suggesting that distributed poisoning reduces the advantage of robust retrieval architectures. At the generation level, openai-gpt-oss-120b remains the more vulnerable model in both settings, since it lowers abstention and increases \gls{ASR}. However, the relative importance of adversarial database coverage is higher in the Micro-Collaborative attack. Overall, the comparison indicates that direct poisoning is more strongly shaped by retriever robustness, whereas Micro-Collaborative Poisoning is more strongly shaped by retrieval depth and multi-source adversarial redundancy.

Taken together, these results show that Micro-Collaborative Poisoning represents a distinct threat from direct poisoning. Direct poisoning relies on strong, highly competitive adversarial documents. Micro-Collaborative Poisoning instead relies on distributed plausibility, redundancy, and accumulation across the retrieved context. This makes the attack particularly relevant for collaborative or multi-source \gls{RAG} environments, where many contributors, documents, or databases may influence the final context. The results suggest that defenses should not focus only on detecting obviously poisoned documents or removing the top adversarial passage. They should also consider weaker signals distributed across multiple passages, the balance between clean and poisoned sources, and the possibility that several individually plausible documents can collectively steer the generator toward an incorrect answer.

%% file: sections/Conclusions.tex
This paper investigated Micro-Collaborative Poisoning as a distributed attack strategy against \gls{RAG} systems, where weak and plausible poisoned content is spread across multiple sources rather than concentrated in a few strongly poisoned documents. The study analyzed how this method of poisoning affects retrieval and generation behavior across different datasets, retrievers, retrieval depths, database compositions, and generator models, using metrics such as Poison@k, Poison Rank, Score Margin, Abstention Rate, \gls{CAR}, and \gls{ASR}.

The results show that Micro-Collaborative Poisoning depends mainly on retrieval breadth and adversarial redundancy across sources. Increasing retrieval depth makes poisoned passages more likely to enter the context, while poisoning multiple databases increases their influence on the generated answer. Clean database diversity improves robustness by strengthening legitimate evidence, whereas adversarial coverage across databases reduces \gls{CAR} and increases \gls{ASR}. Although dense BGE and graph-based retrieval reduce poisoning exposure, their protective effect is weaker than in direct poisoning. Overall, Micro-Collaborative Poisoning can achieve comparable downstream attack success while leaving a weaker document-level poisoning signature, making it harder to detect.

Future work should evaluate the attack across additional datasets, retrievers, and generator models, and investigate defenses tailored to distributed poisoning, such as cross-document consistency checks, source-level trust estimation, redundancy-aware filtering, and coordinated-signal detection. More realistic collaborative settings, where documents evolve over time and contributors have different trust levels, should also be studied to better understand and mitigate this threat in deployed multi-source \gls{RAG} systems.